\documentstyle[12pt]{article}
\date{}
\newcommand{\Bbb}{\bf }

\title{Algorithms for the Computing Determinants in Commutative Rings
       \thanks {This is English version of the paper in:
       Diskretnaya Matematika, {\bf 7}, No.4, 1995, 68--76.}
}
\author{Gennadi~I.~Malaschonok}
\begin{document}
\maketitle

\abstract{
Two known computation methods and one new  computation  method
for matrix determinant over an integral domain are  discussed.  For
each of the methods we evaluate the computation times for different
rings and show that the new method is the best.
}

\section{Introduction}

Among  the  set  of  known  algorithms  for  the   determinant
computation, there is a subset,  which  allows  us  to  carry  out
computations  within  the  commutative  ring   generated   by   the
coefficients of the system. Recently, interest in these  algorithms
grew due to computer algebra computations. These algorithms may  be
used  (a)  to  find  determinant  of  the  matrix  with   numerical
coefficients, (b) to find determinant of the matrix over the  rings
of polynomials with one or many variables over the integers or over
the reals, (c) to find determinant of the matrix over finite fields
and etc.

The first effective  method  for  calculation  of  the  matrix
determinant with numerical coefficients was introduced by   Dodgson
\cite{Dod1866}. Further this method was used in
\cite{WaughDwyer1945} --  \cite{Mal87}. Another method
(one-pass method) was proposed by the author \cite{Mal89}.

Here will be  proposed  more  effective  combined  method  for
calculation of the matrix determinant over an integral domain.

Let  $ A = (a_{ij})$, $i=1 \ldots n, j=1 \ldots n$, be the given
matrix of order $n$
over an integral domain ${\bf R}$.
$$
\delta^{k}=|a_{ij}|, \ \ i,j=1 \ldots k, \ k=1 \ldots n,
$$
denote corner minors of the matrix  $A$ of order $k$, $\delta ^{k}_{ij}$
denotes minors
obtained after a substitution in the minors $\delta ^{k}$ of the column   $ i$ for
the column $j$ of the matrix $A$, $k=1 \ldots n$, $i=1 \ldots k$,
$j=1 \ldots n$.

We examine these three algorithms, assuming  that  all  corner
minors $\delta ^{k}, \ k=1 \ldots n-1$,  of the matrix $ A$ are different
from  zero  and zero divisors. For each of the algorithms we evaluate:

\noindent 1.The  general  computing  time, taking  into  consideration   only
arithmetic operations and assuming moreover, the execution time for
multiplication, division and addition/subtraction of two  operands,
the first of which is a minor of order $ i$ and  the  second  one,  a
minor of order $j$, will be $M_{ij}, D_{ij}, A_{ij}$ correspondingly.

\noindent 2. The exact number of operations of multiplications, division  and
addition/subtrac\-tion over the matrix coefficient.

\noindent 3. The number of operations of  multiplication/division  (M),  when
${\bf R}={\Bbb R}[x_{1} \ldots x_{r}]$ is a ring of polynomials with $r$
variables  with  real
coefficients and only one computer word is required for storing any
one of the coefficients.

\noindent 4. The number of operations of multiplication/divis\-ion $(M_{Z})$,
when ${\bf R}={\Bbb Z}[x_{1} \ldots x_{r}]$ is a ring of polynomials with $r$
variables with  integer coefficients and these coefficient are stored in as
many  computer words as are needed.

\noindent 5. The number of operations of multiplication/divi\-sion $(M_{M})$,
when ${\bf R}={\Bbb Z}[x_{1} \ldots x_{r}]$ is a ring of polynomials with $r$
variables with  integer coefficients, but for computation the modular  method
 is  applied, which is based on the remainder theorem.

\section{Dodgson's algorithm}

Dodgson's algorithm \cite{Dod1866}, \cite{WaughDwyer1945} consists  of $n-1$
steps. In the first step all minors of second order are computed
$$
a^{2}_{ij} = a_{11}a_{ij} - a_{1j}a_{i1}, \  \ i=2 \ldots n, \ j=2 \ldots m,
$$
\noindent which  surround  the  corner  element $a_{11}$. At the $k$-th step,
$k=2 \ldots n-1$, and according to the formula
$$
a^{k+1}_{ij} =
(a^{k}_{kk}a^{k}_{ij} - a^{k}_{ik}a^{k}_{kj})/a^{k-1}_{k-1,k-1},
$$
$$
i=k+1 \ldots n,  \ \ j=k+1 \ldots m,
$$
all the minors $a^{k+1}_{ij}$ of order $k+1$ are computed, which are formed by
surrounding the corner minor $\delta ^{k}$ by row  $ i$ and column $j$, that
is  the minors  which  are  formed  by  the  elements,   located   at   the
intersection of  row $1 \ldots k, i$  and  of  columns $1 \ldots k$, $ j$.
Obviously, $a^{n}_{nj}=\delta ^{n}_{nj}$, $j= n \ldots m$, holds.

Corner minors $\delta ^{k}=a^{k}_{kk},  \ k=1 \ldots n-1,$ must be different
from zero and zero divisors. In order to do so, they can  be  controlled  by
choice of the pivot row or column.

Let us evaluate the computing time of the algorithm
$$
T^{D} = \sum ^{n-1}_{k=1} T^{D}_{k}.
$$
$T^{D}_{1}=(n-1)^{2}(2M_{11}+A_{22})$,
$T^{D}_{k}= (n-i)^{2}(2M_{k,k}+A_{2k,2k}+D_{2k,k-1})$,
$k = 2 \ldots n-1$.
$T^{D}_{k}$ denote the computing time of the $k$ step for Dodgson's algorithm.

\section{One-pass algorithm}
Dodgson's  algorithm  makes  zero  elements  under  the   main
diagonal   of   the   matrix.   One-pass   algorithm   \cite{Mal89}    makes
diagonalisation of the coefficient matrix minor-by-minor and  step-by-step.

This algorithm consists of $n-1$ steps. In the  first  step  the
minors of the second order are computed
$$
\delta ^{2}_{2j} = a_{11}a_{2j} - a_{21}a_{1j}, \ j=2 \ldots n,
\delta ^{2}_{1j} = a_{1j}a_{22} - a_{2j}a_{12}, \ j=3 \ldots n.
$$
In the $k$-th step, $k=2 \ldots n-1,$  the  minors  of  order $k+1$  are
computed according to the formulae
$$
\delta ^{k+1}_{k+1,j} = a_{k+1,k+1}\delta ^{k}_{kk} -
\sum ^{k}_{p=1} a_{k+1,p}\delta ^{k}_{pj}, \ j=k+1 \ldots n,
$$
$$
\delta ^{k+1}_{ij} = {\delta ^{k+1}_{k+1,k+1}\delta ^{k}_{i,j} -
\delta ^{k+1}_{k+1,j}\delta ^{k}_{i,k+1}\over  \delta ^{k}_{k,k}}, \
i=1 \ldots k, \ j=k+2 \ldots n.
$$
In this way, at the $k$-th step the coefficients  of  the  first
$k+1$ rows  of  the  matrix  take  part. Corner  minors $\delta ^{k}$ can be
controlled by the choice of the pivot row or column.

The general computing time of the one-pass algorithm is
$$
T^{D}_{1}=(2n-3)(2M_{1,1}+A_{2,2}),
$$
$
T^{D}_{k}=(n-k)((k+1)M_{k,1}+kA_{k+1,k+1})+k(n-k-1)
(2M_{k,k+1}+A_{2k+1,2k+1}+D_{2k+1,k}),$  \
$k = 2, \ldots, n-1, \ \
T^{D} = \sum ^{n-1}_{k=1} T^{D}_{k}.
$

\noindent $T^{O}_{k}$ denote the computing time of the $k$ step for one-pass algorithm.

\section{Combined algorithm}
We can get more effective algorithm if we  will  combine  one-pass
algorithm (it will be first part) and Dodgson's algorithm  (it
will be the second part).

We shell make diagonalisation of the first part  (upper  part)
of the matrix and then we shall make zero elements under  the  main
diagonal of the second part (lower part) of the matrix.

In the first part we will execute $r-1$ steps  of  the  one-pass
algorithm. In the first step the minors of order 2 are computed
$$
\delta ^{2}_{2j} = a_{11}a_{2j} - a_{21}a_{1j}, \  j=2 \ldots n,
$$
$$
\delta ^{2}_{1j} = a_{1j}a_{22} - a_{2j}a_{12}, \ j=3 \ldots n.
$$
In the $k$-th step, $k=2 \ldots r-1,$ the minors of order $k+1$ are computed
$$
\delta ^{k+1}_{k+1,j} = a_{k+1,k+1}\delta ^{k}_{kk} -\sum ^{k}_{p=1} a_{k+1,p}\delta ^{k}_{pj}, j=k+1 \ldots n,
$$
$$
\delta^{k+1}_{ij} =
{ \delta^{k+1}_{k+1,k+1}\delta^{k}_{i,j} -
\delta^{k+1}_{k+1,j}\delta^{k}_{i,k+1} \over \delta^{k}_{k,k} },
\ i=1, \ldots k, \ j=k+2 \ldots n.
$$
Then, in the $r$ step, we can compute all  minors $a^{r+1}_{ij}$  of  the
order $r+1,$ which are formed by surrounding the corner  minor
$\delta ^{r}_{rr}$ of order $r$ by row $ i$ and column $j \ (i>r, j>r)$
$$
\delta ^{r+1}_{i,j} =  a_{i,r+1}\delta ^{r}_{rr} -
\sum ^{r}_{p=1} a_{i,p} \delta ^{r}_{pj}, i,j=r+1, \ldots, n.
$$
In the second  part  we  will  execute  last $n-r-1$  steps  of
Dodgson's algorithm according to the formula
$$
a^{k+1}_{ij} = {a^{k}_{kk}a^{k}_{ij} - a^{k}_{ik}a^{k}_{kj}\over
a^{k-1}_{k-1,k-1}  }, \
k=r+2 \ldots n-1, \ i,j=k+1, \ldots, n.
$$
Obviously, $a^{k}_{kk} = \delta ^{k}$, $ k=2 \ldots n$, holds and
$a^{k}_{kk}$  is  the  matrix determinant.

Here we have $n-3$ different variants of the combined algorithm,
because $r$ may  be  equal  to $2,3, \ldots ,n-2.$  We  will  have  one-pass
algorithm if $r$ will be equal $n-1.$

The computing time of this algorithm is
$$
T = \sum ^{n-1}_{k=1} T_{k}
$$
$
T_{r}= (n-r)^{2}((r+1)M_{r,1}+rA_{r+1,r+1}),
$
$T_{k}=T^{O}_{k}$ for $k=1 \ldots r-1$. $T_{k}=T^{D}_{k}$ for
$k=r+1 \ldots n-1$.
$T^{O}_{k}$ and $T^{D}_{k}$ denote the computing time of  the $k$-th
step  for  one-pass algorithm and Dodgson's algorithm correspondingly.

\section{Evaluation of the quantity of operations \\ over the matrix elements}

We have now $n-1$ different  methods,  if  Dodgson's  method  is
considered as one of them. And we will evaluate  the  calculation
time  for  each method.

We begin the comparison  of  the  algorithms  considering  the
general  number   of   multiplications $N^{m}$,   divisions $N^{d}$   and
additions/subtractions $N^{a}$, which are necessary for  calculation  of
the matrix determinant. Moreover, we will not make any  assumptions
regarding the computational complexity of these operations; that is
we  will  consider  that  during  the  execution   of   the   whole
computational process, all multiplications of the coefficients  are
the   same,   as   are   the   same   all   divisions    and    all
additions/subtractions.

The quantity of operations, necessary for  Combined  algorithm
with arbitrary $r$ will be \\
$N^{r}_{a} = (2n^{3}-3n^{2}+n)/6,$ \\
$N^{r}_{m} = (4n^{3}-4n-4r^{3}+9r^{2}n-6rn^{2}-3rn+4r)/6,$ \\
$N^{r}_{d} = (2n^{3}-3n^{2}-5n+12-4r^{3}+9r^{2}n-3r^{2}-6rn^{2}+3rn+r)/6.
$

It is easy to see, the most  effective  algorithm  is  combined
algorithm with $r=n/2$ if $n$ is odd and $r=(n+1)/2$ if $n$ is even $(n>2)$.

Then we can compare all three algorithms

\bigskip
\noindent
\begin{tabular}{|c|c|c|}
\hline
\multicolumn{3}{|c|}{Quantity of operations} \\
\hline
algorithm & $N^m$  &  $N^d$ \\
\hline
Dodgson's & $(4n^3-6n^2+2n)/6)$ &  $(2n^3-9n^2+13n-6 )/6$ \\
\hline
One-pass  & $(3n^3-3n^2)/6$     &  $(n^3-3n^2-4n+12)/6$ \\
\hline
Combined,  & $(11n^3-6n^2-(8+3v )n +6v)/24$
&  $(3n^3-9n^2-(18-3v )n+$ \\
$r=(n+v)/2$ &  &  $+48-3v ) / 24$ \\
\hline
\end{tabular}

\bigskip
\noindent
$v=0$ if $n$ is odd and $v=1$ if $n$ is even $(n>2)$. The quantity of the
additions/subtractions $(N^{a})$ operations is the same for these three
algorithms
$$
N^{D}_{a} = N^{O}_{a} = N^{C}_{a} = (2n^{3}-3n^{2}+n)/6.
$$
If we evaluate quantity of operations,  considering  only  the
third power, then we obtain the evaluation
$N^{D}_{m}:N^{O}_{m}:N^{C}_{m} = 16:12:11$, $N^{D}_{d}:N^{O}_{d}:N^{C}_{d}
= 8:4:3.$

If we evaluate according to the general quantity of  multiplication and 
division operations, considering only the  third  power,
then we obtain the evaluation $12n^{3}/12 : 8n^{3}/12 : 7n^{3}/12.$

So, according to this evaluation, combined algorithm is to  be
preferred.

\section{Evaluation of the algorithms in the ring \\
$\bf R[x_1 \ldots x_s]$}

Let ${\bf R}$ be the ring  of  polynomials  of $s$  variables  over  an
integral domain and let us suppose that every element $a_{ij}$  of  the
matrix  $ A$  is a polynomial of degree $p$ in each variable
$$
a_{ij}= \sum^{p}_{u=0}\sum ^{p}_{v=0}\ldots
\sum ^{p}_{w=0} a_{uv \ldots w}x^{u}_{1}x^{v}_{2} \ldots x^{w}_{r}.
$$
Then it is possible to define, how much time is  required  for  the
execution of the arithmetic operations over polynomials  which  are
minors of order  $ i$ and $j$ of the matrix  $ A$ \\
$
A_{ij}=(jp+1)^{s}a_{ij}, \\
$
$
M_{ij}=(ip+1)^{s}(jp+1)^{s}(m_{ij}+a_{i+j,i+j}), \\
$
$
D_{ij}=(ip-jp+1)^{s}(d_{ij}+(jp+1)^{s}(m_{i-j,j}+a_{ii})). \\
$
Here we  assume,  that  the  classical  algorithms  for  polynomial
multiplication and division are used. And besides, we consider that
the time necessary for execution of the  arithmetic  operations  of
the coefficients of the  polynomials   is $m_{ij}, d_{ij}, a_{ij}$,  for
the operations of multiplication,  division  and  addition/subtraction,
respectively,  when  the  first  operand  is  coefficient  of   the
polynomial, which is a minor of order  $ i$, and the second -- of  order
$j$.

Let us evaluate  the  computing  time  for  each  of  the $n-1$
algorithms, considering that the coefficients  of  the  polynomials
are real numbers and each one is stored in one  computer  word.  We
will   assume   that $a_{ij}=0, m_{ij}=d_{ij}=1, A_{ij}=0$,
$M_{ij}=i^{s}j^{s}p^{2s}, D_{ij}=(i-j)^{s}j^{s}p^{2s}$,
$\sum ^{n-1}_{k=1} i^{p} = n^{p+1}/(p+1) - n^{p}/2 + O(n^{p-1})$,
and we will consider only the leading terms in $n$ and $r$:
$$
M(r)=3p^{2s} \bigg(
{2n^{2s+3} \over (2s+1)(2s+2)(2s+3)} 
$$
$$
- {r^{2s} \over 2} \bigg({4r^3 \over 2s+3} - 6r^{2} {n+s+1\over 2s+2} +
nr{ 2n+12s+7 \over 2s+1} -n^{2} \bigg) \bigg)
$$
We have
$M^{D}= M(r)$ for $r=0$, $M^{O}= M(r)$ for $r=n$,
$ M^{O}= M^{D}(2s+1)/2. $
$M^{O}$ and $M^{D}$ denote the computing  time  for  one-pass  algorithm  and
Dodgson's algorithm.

It is easy to see, the most  effective  algorithm  is  combined
algorithm with $r_{best}=n/2 - 3s/2 + 2 + O(n^{-1})$.
For $n,r>>s>1$ we obtain $r_{best}=n/2$.

\section{ Evaluation \ of the algorithm \ in the ring \\
${\Bbb Z} \bf [x_1 \ldots x_s]$,  \  standard case}

As before we suppose that every coefficient of the matrix is a
polynomial. However, the coefficients of these polynomials are  now
integers and each one of these coefficients $a^{ij}_{uv \ldots w}$
is stored  in $l$
computer words. Then, the coefficients of the polynomial, which  is
a minor of order $ i$, are integers of length $i l$ of computing words.

Under the assumption that classical algorithms  are  used  for
the arithmetic  operations  on  these  long  integers,  we  obtain:
$a_{i j}=2j l a$, $m_{ij}=i j l^{2}(m+2a)$,
$d_{i j} =(i l-j l+1)(d+j l(m+2a))$, where $a, m, d$ -
are the  execution  time  of  the  single-precision  operations  of
addition/subtraction, multiplication, and division.

Assuming that $a=0, m=d=1,$ we obtain the  following  evaluation
of the execution times of polynomial  operations:
$M_{i j}=i j l^{2}(i j p^{2})^s$,
$D_{i j}=(i-j)^{s+1} j^{s+1} l^{2} p^{2s}$, $A_{i j}=0.$

In this way, the evaluation of the computing time will be  the
same as that for the ring
${\bf R}={\Bbb R}[x_{1},x_{2} \ldots x_{s}]$, if we replace  everywhere
$s$ by $s+1$ and $p^{s}$ by $lp^{s}$.

Therefore, the most effective algorithm is combined  algorithm with
$ r_{best}=n/2 - 3s/2 + 1/2 + O(n^{-1}).$
For  $n,r>>s>1$ we obtain  $r_{best}=n/2$.

\section{Evaluation of the algorithms in the ring \ \ \\ 
${\Bbb Z}  \bf [x_1 \ldots x_s]$, \ 
modular case}

Let us evaluate the time for the solution of the same problem,
for  the  ring  of  polynomials  with $s$  variables  with   integer
coefficients ${\bf R}={\Bbb Z}[x_{1} \ldots x_{s}]$, when the  modular
method  is  applied  --
based on the remainder theorem. In this case we will not take  into
consideration the operations for transforming the  problem  in  the
modular form and back again.

It suffices to define the number of moduli,  since  the  exact
quantity of operations on the matrix elements for  the  case  of  a
finite field has already been obtained in section 5.

We will consider that every prime  modulus $m_{i}$  is  stored  in
exactly one computer  word,  so  that,  in  order  to  be  able  to
recapture the polynomial coefficients, which are minors of order
$n$, the $n(l+\log (n p^3)/2 \log m_i)$ moduli are needed,
what is easy to see  due  to Hadamar's inequality.

Further, we need up moduli for each unknown $x_{j}$, which  appears
with maximal degree  $ n p$. There are $s$ such unknowns,  and  therefore,
in all, $\mu =p s n^2(l+\log (np^3) /2 \log m_{i})$ moduli are needed.

If we now make use of the table in section 4, denote the  time
for modular multiplication by $m$ and the time for  modular  division
by $d$, then not  considering  addition/subtraction  and  considering
only leading  terms  in $n$,  we  obtain :
$$
M^{D}_{M}=(16m+8d)\nu, \ \  M^{O}_{M}=(12m+4d)\nu, \ \
M^{C}_{M}=(11m+3d)\nu, \ \ \hbox{ where } \nu  =\mu n^{3}/3.
$$

\section{Conclusion}

We see, Dodgson's method is better than  the  one-pass  method
for non-modular  computation  in  polynomial  rings,  and  one-pass
method is better than Dodgson's method in other cases, but combined
method with $r=n/2$ is the best in all cases.

%%%%%%% BIBLIO %%%%%%%

\bigskip\noindent\bibliographystyle{plain}

\end{document}